\documentclass[reprint, amsmath,amssymb,
aps,showkeys,
pre]{revtex4-1}

\usepackage{amssymb}
\usepackage{amsmath}

\usepackage{graphicx}
\usepackage{graphics}
\usepackage{dcolumn}
\usepackage{color}
\usepackage{rotate}
\usepackage{fancyhdr}
\usepackage{hyperref}
\usepackage{indentfirst}

\usepackage{umoline}
\usepackage{ulem}

\usepackage{dcolumn}
\usepackage{bm}

\usepackage[dvipsnames]{xcolor}

\usepackage{setspace}
\usepackage{color}
\usepackage{float}

\newcommand{\be}{\begin{equation}}
\newcommand{\ee}{\end{equation}}

\newcommand{\bea}{\begin{eqnarray}}
\newcommand{\eea}{\end{eqnarray}}

\renewcommand{\Im}{{\rm \, Im\,}}

\begin{document}
	
	\title{Second harmonic generation as a minimal model of turbulence}
	\date\today
	\author{N. Vladimirova$^{1,2}$}
	\author{M. Shavit$^{2}$ }
	\author{S. Belan$^{3,4}$}
	\author{G. Falkovich$^{2}$ }
	\affiliation{$^{1}$ Brown University,  Providence, RI 02912, USA}
	\affiliation{$^{2}$ Weizmann Institute of Science, Rehovot 76100 Israel}
	\affiliation{$^{3}$ Landau Institute for Theoretical Physics, 
		142432 Chernogolovka, Russia}
	\affiliation{$^{4}$
		National Research University Higher School of Economics, 
		101000 Moscow, Russia}
	
	\begin{abstract}
		When two resonantly interacting modes are in contact with a thermostat, their statistics is exactly Gaussian and the modes are statistically independent despite strong interaction. Considering noise-driven system, we show that when one mode is pumped and another dissipates, the statistics (of such cascades) is never close to Gaussian no matter the interaction/noise relation. One finds substantial phase correlation in the limit of strong interaction (weak noise). Surprisingly,  for both cascades, 
	the mutual information between modes increases and entropy further decreases when interaction strength decreases.  We use the model to elucidate the fundamental problem of far-from equilibrium physics: where the information (entropy deficit) is encoded and how singular measures form. For an instability-driven system (a laser), even a small added noise leads to large fluctuations of the relative phase near the stability threshold, while  far from it we show that the conversion into the second harmonic is weakly affected by noise.\end{abstract}
	\maketitle

\section{Introduction}
\label{sec:1}

Second harmonic generation is the simplest fundamental process of nonlinear wave physics, which is also in the center of numerous practical applications in laser physics and beyond. The dynamics of the process has been studied exhaustively \cite{SG1,SG2}, which cannot be said about statistics, despite the fact that understanding the influence of noise on the energy conversion is of paramount practical importance, recently enhanced by the use of meta-materials \cite{SGH1}. Here we address this problem by studying theoretically a two-mode resonant system driven by a combination of pumping and random  noise.  Our motivation is two-fold. Apart from the classical conversion problem, we find this system ideally suited for elucidating the fundamental problems of non-equilibrium physics. When one mode is stochastically forced  and another is dissipated, that presents a minimal model of turbulence cascade. The freedom to force either mode allows us to elucidate the basic differences between direct and inverse cascades. Apart from energy, we shall be interested in the entropy of such far-from equilibrium state, which is expected to be much lower than in thermal equilibrium with the same energy.

A remarkable property, common for all systems of resonantly interacting waves and shared with hydrodynamic systems \cite{VSF}, is that the canonical thermal equilibrium has exactly Gaussian statistics, and the modes fluctuate independently, regardless of the interaction strength. 
Here we describe how deviations from equilibrium diminish entropy and build correlations between the two modes. Far from equilibrium the joint two-mode  statistics is never close to Gaussian, even when the marginal distribution of every mode is close to Gaussian. On the one hand, the entropy decrease means that the statistical distribution is getting more non-uniform, which poses the question: Can it lead all the way to singularity like the celebrated Sinai-Ruelle-Bowen measures in dynamical systems \cite{D,FS}? We show that this is indeed so: the measure in the phase space is getting singular in the double limit of strong non-equilibrium and weak interaction. On the other hand, since entropy is missing information, any entropy decrease poses another question: Where all this extra information about non-equilibrium is encoded? First, we find out how the entropies of the three marginal distributions, of each mode amplitude and their phase difference, go down as the system deviates from equilibrium. Second, we find out which part of the entropy decrease is due to inter-mode correlation. This is properly measured by the mutual information (rather than by the pair correlation function, suitable for Gaussian statistics only).

The process of the second harmonic generation is described by the following model Hamiltonian (assuming perfect resonance)
\begin{align} {\cal H}_0 =\omega  |a_1|^2  +2\omega   |a_2|^2 +   V  a_1^{*2}a_2+  V^*a_1^{ 2}a_2^*  \ .\label{Ham5}\end{align}
Here  $a_1$ and $a_2$ are the complex amplitudes of two non-linearly coupled modes having frequencies  $\omega$ and $2\omega$, respectively, and $V$ is the interaction constant (considered real positive without loss of generality).
The two coupled complex equations govern  dynamics: $\dot a_k=-i {\partial {\cal H}_0}/{\partial a_k^*}$, $k=1,2$.
We eliminate the linear  terms in these equations by introducing the envelopes
\begin{equation}\label{env}
b_1=a_1e^{i\omega t},  b_2=a_2e^{2i\omega t}.
\end{equation}
That results in a strongly interacting system with a cubic Hamiltonian ${\cal H} =  V  b_1^{*2}b_2+  V^*b_1^{ 2}b_2^* $.

Due to the symmetry $b_1\to b_1e^{i\phi}$, $b_2\to b_2e^{2i\phi}$ the system $\dot b_k=-i {\partial {\cal H}}/{\partial b_k^*}$ has an extra integral of motion $N=  |b_1|^2  +2  |b_2|^2$ and is completely integrable; the phase portrait is presented in Appendix~\ref{sec:Ham}.
Let us add dissipation and stochastic pumping:
\bea
\label{gen00}
&&\dot b_1=-2i V^*b_1^*b_2-\gamma_1b_1+ \xi_1(t),\\
&&\dot b_2=-i Vb_1^2-\gamma_2 b_2+ \xi_2(t).
\label{gen0}\eea
Here $\gamma_1$ and $\gamma_2$ are the damping coefficients, and $\xi_1$ and $\xi_2$ are  independent Gaussian random forces with zero mean $\langle \xi_i(t)\rangle=0$ and the variance $\langle \xi_i(t_1) \xi_j^*(t_2)\rangle=P_i\delta_{ij}\delta(t_2-t_1) $.


We mainly focus on the properties of the statistically steady solutions of the system (\ref{gen00},\ref{gen0}) in the case when one mode is forced, while the other is damped. 
Since the modes enter the Hamiltonian in a non-symmetric way, there are two possibilities: one either can pump the first (lower frequency) mode and dump the second (higher frequency) mode or vice versa.
The former scenarios qualitatively corresponds to the direct energy cascade, while the second is reminiscent to the inverse cascade.

\begin{figure}[h!] 
	\includegraphics[width=50mm]{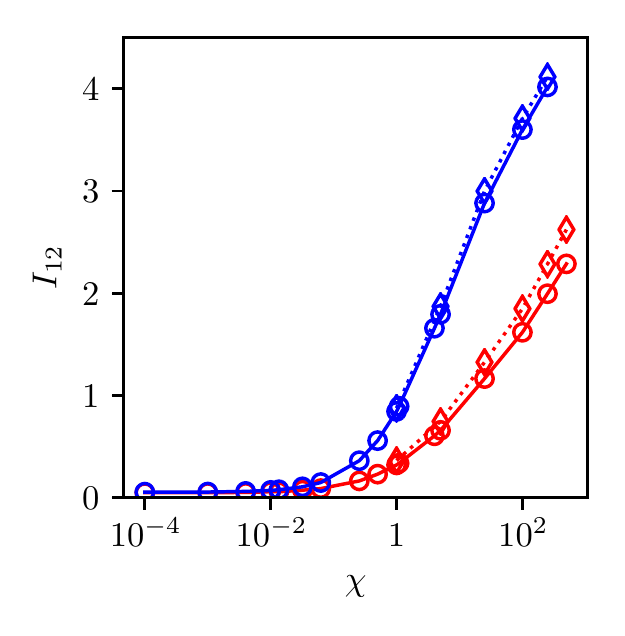} \centering
	\caption{
		Mutual information versus Reynolds number for direct and inverse cascades (red and blue lines respectively). Three dimensional
		distribution are computed with bin size $\Delta \theta = 2\pi/32$ and $\Delta\rho^2_{1,2}/n_{1,2} = 1$ (circles) and 0.5 (diamonds).
	}
	\label{tm_mi_ext}
\end{figure}

We wish to understand how much information is needed in order to build a turbulent state and how much one learns about one mode by observing another. For that we will use the metrics from information theory: entropy and mutual information. The answer to the first question is given by the decrease in entropies
\bea &S_{12}=-\int db_1db_1^* db_2db_2^*\,\rho(b_1,b_2)\ln\rho(b_1,b_2)\,,\nonumber \\&S_1=-\int db_1db_1^*\, \rho(b_1 )\ln \rho(b_1 )\,,\nonumber \\&S_{ 2}=-\int   db_2db_2^*\,\rho( b_2)\ln\rho( b_2)\,,\nonumber\eea
where $\rho$ is either full or marginal probability distribution. The answer to the second question is given by the mutual information between the modes:
\be I_{12}=S_1+S_2-S_{12}\ .\label{MI}\ee
Fig.\ref{tm_mi_ext} demonstrates the growth of the mutual information versus the degree of non-equilibrium (an analog of the Reynolds number  defined below, see (\ref{chi})).



As one of the simplest model of energy transfer, the system of two coupled oscillators  has received considerable attention in the literature~\cite{Kumar_2008,Bonetto_2004,Ciliberto_2013,Falasco_2015,Chun_2015,
	Mura_2018,Horowitz_2019,Courant1,Courant}. In particular, in the mathematical literature, one finds an analysis of a two-mode system with a quadratic Hamiltonian ${\cal H} = T  a_1^{*2}a_2^2$ with the purpose to get insight into the energy transfer in wave turbulence \cite{Courant1,Courant}. 
What distinguishes our model is that it directly corresponds to physical reality and allows experimental validation. In addition, an asymmetry between the modes allows us to compare direct and inverse cascades, which turn out quite different. Another distinction is that we add entropic and informational consideration to the energetic analysis.

The paper is organized from the point of view of entropy: we start from the maximal-entropy equilibrium and investigate near-equilibrium states in Section~\ref{eq}. We then move to study the noise-driven direct and inverse cascades in Section \ref{sec:turb}. We define a dimensionless parameter akin to the Reynolds number and show how entropy decreases as the Reynolds number increases. We begin with the limit of small Reynolds number in \ref{smallRe} and continue to large Reynolds number in \ref{sec:largeRe},  where entropy decreases all the way down as the measure becomes singular in the limit of infinite Reynolds number. In Section~\ref{sec:laser} we consider an instability-driven first harmonic  and study the influence of noise on the conversion process, this can serve as a simple model for a laser generating second harmonic. Conclusion~\ref{sec:con} briefly lists our main results.


\section{Near Thermal equilibrium}
\label{eq}

Adopting the language of stochastic thermodynamics, one can call the ratios $P_1/\gamma_1\equiv T_1$ and $2P_2/\gamma_2\equiv T_2$  effective temperatures experienced by two modes $b_1$ and $b_2$ which are governed by the Langevin equations (\ref{gen00}) and (\ref{gen0}).
If $\Delta T=T_1-T_2=0$, then it is straightforward to find from the Fokker-Planck equation or entropy maximum the steady-state joint probability distribution:
\be {\cal P}_0={1\over Z} \exp\biggl(-\frac{2|b_1|^2+4|b_2|^2}{T}\biggr)={1\over Z} \exp\biggl(-\frac{2N}{T}\biggr).\label{SG2}\ee
Despite strong interaction,  this distribution is exactly Gaussian and the modes are statistically independent.
The later means that the mean energy flux and the mutual information between modes are zero. Thermal equilibrium corresponds to the equipartition of the quadratic invariant: $\langle |b_1|^2\rangle\equiv n_1=2n_2\equiv 2\langle |b_2|^2\rangle$.

What can we say about the system's statistics when modes are subject to  different effective temperatures?
Let us introduce the dimensionless measure of non-equilibrium
\begin{equation}
\sigma=\frac{\Delta T}{T},\ \  \text{where} \ \ T=\frac{P_1+2P_2}{2(\gamma_1+\gamma_2)}.
\end{equation}
Another  dimensionless parameter quantifies interaction strength relative to the dissipation:
\be\chi={(\gamma_1+\gamma_2)^3\over (P_1+2P_2)|V|^2}\ .\label{chi}\ee

\begin{figure*} 
	\includegraphics[width=175mm]{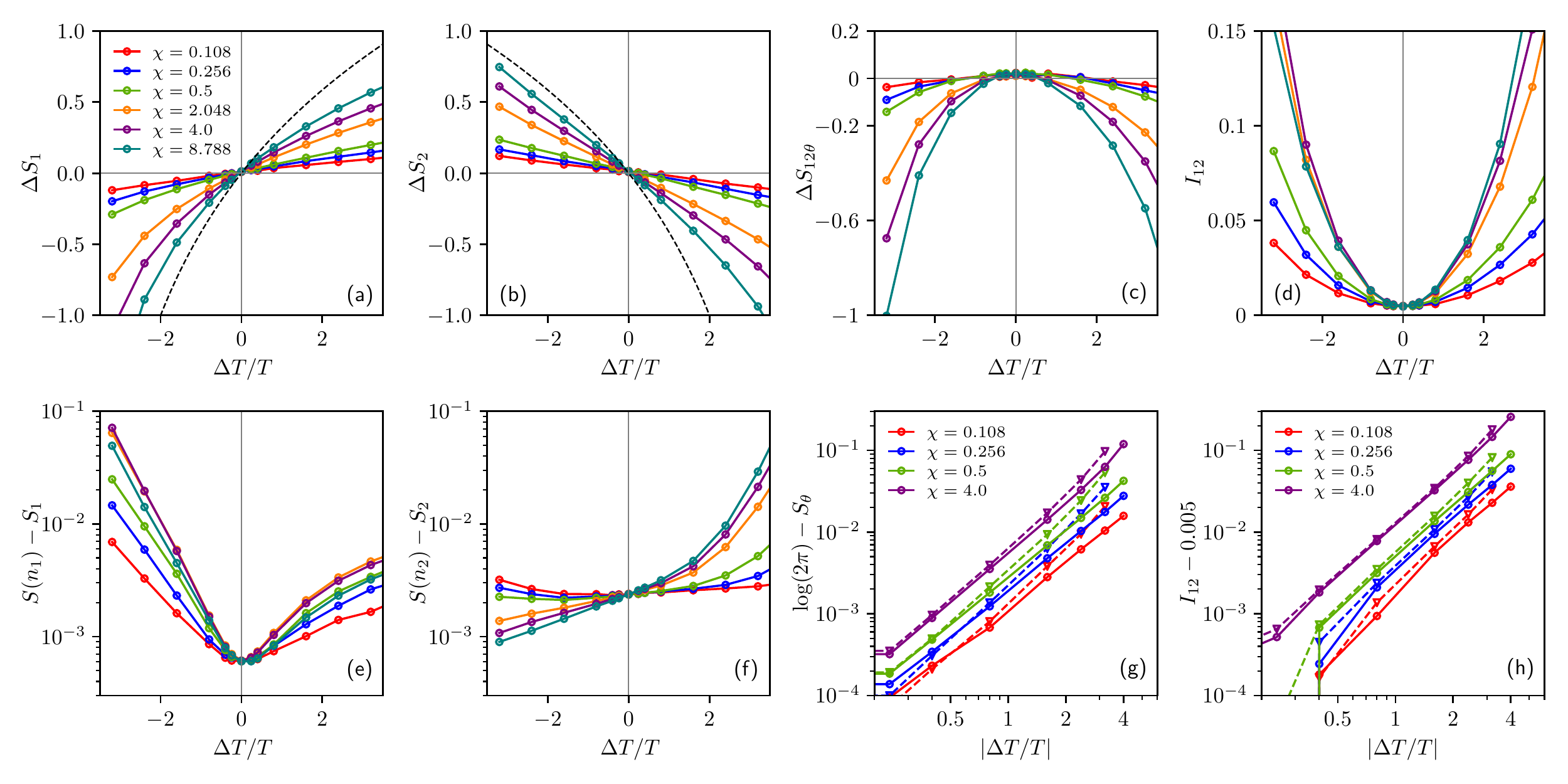}\centering
	\caption{
		Top row: deviation of entropies from equilibrium ($\Delta T = 0$) and mutual information
	for $\gamma_1 = \gamma_2$  and different strength of interaction marked by color.
		 Equilibrium is Gaussian with
		$S_{{\rm eq}1} = 1/\ln(2)$,
		$S_{{\rm eq}2} = 1/\ln(2) - 1$, 
		and
		$S_{{\rm eq}\theta} = \log_2(2 \pi)$.
		Dashed lines show the entropy deviation of a single mode respectively at $T_1$ and $T_2$
		for marginal distributions,
		$\Delta S_1 =  \log_2 (1 + \Delta T/4T) $ and
		$\Delta S_2 = \log_2 (1 - \Delta T/4T)$ .
		The entropies of individual modes are affected by the change of variance of (nearly Gaussian)
		distribution and by deviation from the Gaussian shape. For close-to-equilibrium cases presented here
		the first effect is dominant.
		Bottom row:  panels (e) and (f) show  deviation of entropies from the entropies of Gaussian distributions with the same variance;
		panel (g) illustrates the quadratic dependence of the entropy of phase distribution,
		panel (h) illustrates the quadratic dependence for mutual information
		(solid lines  for $\Delta T > 0$; dashed lines for  $\Delta T <0$). The offset 0.005 is the effect of
		 finite ensemble and bin size; the offset decreases as the size of data set and resolution improve.
	3D distributions are computed with bin sizes $2\pi/32$ for the phase and
		$0.1 T$ for $\rho_{1,2}$.
	}
	\label{fig_s12}
	\label{fig_mi}
\end{figure*} 

Denote $\rho_{1,2}=|b_{1,2}|$ and $\theta=\arg(b_1^2b_2^*)$.  From (\ref{gen00}) and (\ref{gen0}), the steady-state equations on the second moments read:
\begin{eqnarray}
\label{moments1}
&&-4V\langle \rho_1^2\rho_2\sin\theta\rangle -2\gamma_1\langle\rho_1^2\rangle+P_1=0,\\
\label{moments2}
&&2V\langle \rho_1^2\rho_2\sin\theta\rangle -2\gamma_2\langle\rho_2^2\rangle+P_2=0,
\end{eqnarray}
The time derivative of the real part of the third moment is given by: $d\langle {\cal H}\rangle/dt=-(2\gamma_1+\gamma_2)\langle \cal H\rangle$, since $\langle \xi_1b_1b_2^*\rangle=\langle \xi_2^*b_1^2\rangle=0$. Therefore, in any steady state, either in thermal equilibrium or out of it, one has
\begin{equation}
\label{moments3}\langle {\cal H}\rangle=2V\langle\rho_1^2\rho_2\cos\theta\rangle=0.
\end{equation}
Equations (\ref{moments1})-(\ref{moments3}) are valid for any values of $\sigma$ and $\chi$.

{At $\sigma\ne0$, the  probability density ${\cal P}(b_1,b_2)$ is non-Gaussian in non-equilibrium, yet it is close to Gaussian when $|\sigma|\ll1$  for all values of $\chi$.
	The simplest to treat is the limit of small interaction, $\chi \gg1$.
	In this case, the first correction to Eq. (\ref{SG2}) is determined by the energy flux between modes, which is small and proportional to the temperature difference:}
\begin{eqnarray}
\label{pert4}
&&\ln{\cal P}(b_1,b_2) \approx - { 2|b_1|^2\over T_1}-{ 4 |b_2|^2 \over  T_2}-\\
&&-{4\Delta T\over P_1T_2+P_2T_1}\Im [V^*b_1^{*2}b_2]+O(\chi^{-2})
\end{eqnarray}

Smallness of interaction multiplies the parameter of  non-equilibrium  $\Delta T/T$ in the right hand side of Eq. (\ref{pert4}), so that this result is valid  even when $\Delta T/T$ is not small.
That means that, as long as both temperatures remain finite and interaction is weak, even far from equilibrium  the relative entropy is small:
$$D({\cal P}|{\cal P}_0)=\int db_1db_1^*db_2db_2^*{\cal P}\ln({\cal P}/{\cal P}_0)\propto \chi^{-1}\ll1\ ,$$
as well as the mutual information.

In the opposite limit $\chi\ll1$ or $V\to\infty$, the non-Gaussian correction is again proportional to the product of the degree of non-equilibrium and the small parameter $\chi$. In terms of $x=|b_1|^2$ and  $y=2|b_2|^2$ we obtain:
$$\ln{\cal P}  \approx- {x +y \over T}+{\chi^{1/2}\Delta T\over  T  }f(x,y,\theta)\ ,$$
where the correction satisfies the equation
$${2x {y}\over\sqrt{yT}}\left[\sin\theta\left({\partial\over\partial y}-{\partial\over\partial x}\right)
+{x-2y\over2xy} \cos\theta{\partial\over\partial \theta}\right]f =x-y\,.$$
The correction is odd in phase difference, $f(-\theta)=-f(\theta)$, and scales linearly with amplitudes, so that it is substantial at small amplitudes.  
In the limits,  $f\to g(\theta)\sqrt{y/2T}$, where $g=\sin\theta$ at $y\ll x$, and  $g=\int d\theta/\cos\theta$ at $y\gg x$.

It makes sense to compare entropies at the same mean quadratic energy $N$. To see how entropy goes down on the way to turbulence we shall subtract  the total entropy from its maximal equilibrium value, which quantifies the amount of information one needs to create a turbulent state:   $\Delta S(N,n_2/n_1)=S_0- S_{12}$.

Numerics support quadratic decrease of $S_{12}(\Delta/T)$ and increase of $I_{12}(\Delta/T)$ up to $\Delta T\simeq 4T$, see Figure~\ref{fig_s12}. 

When $\Delta T/T$ exceeds one, the functions are not even which demonstrates the statistical difference between upward and downward energy conversion. We see stronger deviations from Gaussianity for negative $\Delta T<0$, which corresponds to the downward energy flow and to an inverse cascade at $\Delta T/T \to- \infty$. The physical difference  is that the first mode pumps the second one as an additive force, while the second mode pumps the first one as a multiplicative instability. Therefore, it seems natural  that the entropy is generally lower and the mutual information higher for an inverse transfer.  The analysis of the separate distributions of two amplitudes and the relative phase shows that the entropy of the driven mode (say, $S_1(\Delta/T)$ for a direct transfer) grows with $\Delta T$ slower than the entropy of the dissipated mode and $S_\theta(\Delta/T)$ decrease, see panels (g) and (h) in Figure \ref{fig_mi}.



\section{Turbulent cascades}\label{sec:turb}

Now let us have an energy cascade in our model: pumping one mode and dissipating another. When energy flows from lower frequency mode to higher, i.e $0=\xi_2=\gamma_1$ in Eqs. (\ref{gen00}) and (\ref{gen0}), the cascade is called direct, and inverse when $0=\xi_1=\gamma_2$.
In these cases, the only dimensionless parameter is $\chi=\gamma^3/(P|V|^2)$, where $P$ is the intensity of noise acting upon the driving mode, and $\gamma$ denotes the damping coefficient of the dissipating mode. As we shall see below, $\chi$ to some extent plays the role of the Reynolds number of hydrodynamics in a sense that it determines how low is the entropy and how much the occupation numbers deviate from the equipartition $n_1=2n_2$, 
even though the system is not close to thermal equilibrium for however small or large $\chi$.

Balance of the quadratic invariant, $N$, means that the dissipating mode keeps the magnitude of order of its equilibrium value: $n_2= P/4\gamma$ for the direct cascade and $n_1=P/\gamma$ for the inverse cascade. How much the mode which is pumped exceeds the equipartition value is determined by the value of $\chi$, as described below. Note that this  parameter can be interpreted as the squared ratio of the dissipation rate $\gamma$ and the nonlinear transfer rate $Vn\simeq V\sqrt{P/\gamma}$.

When $\chi$ is small, the interaction between modes is strong and the energy transfer is fast, so that  the occupation numbers are expected to be close to equipartition, yet the statistics is not expected to be close to  separable Gaussian form given by (\ref{SG2}). Even though the noise is weak, it is white, that is a singular perturbation destroying integrability everywhere in the phase space \cite{Kurchan}; we shall see below how non-trivial the probability distribution is already in this limit.

One may naively expect that in the opposite limit of large $\chi$, when the noise is strong and interaction is weak, the correlation between modes would be weak too. We shall show below that the opposite is true far from equilibrium: the necessity to carry the flux makes the modes strongly correlated precisely because of a strong noise and weak interaction. It is in this limit we find the lowest entropy and the maximal mutual information between modes, as well as appearance of singular measure in phase space.

\begin{figure*}. 
	\includegraphics[width=175mm]{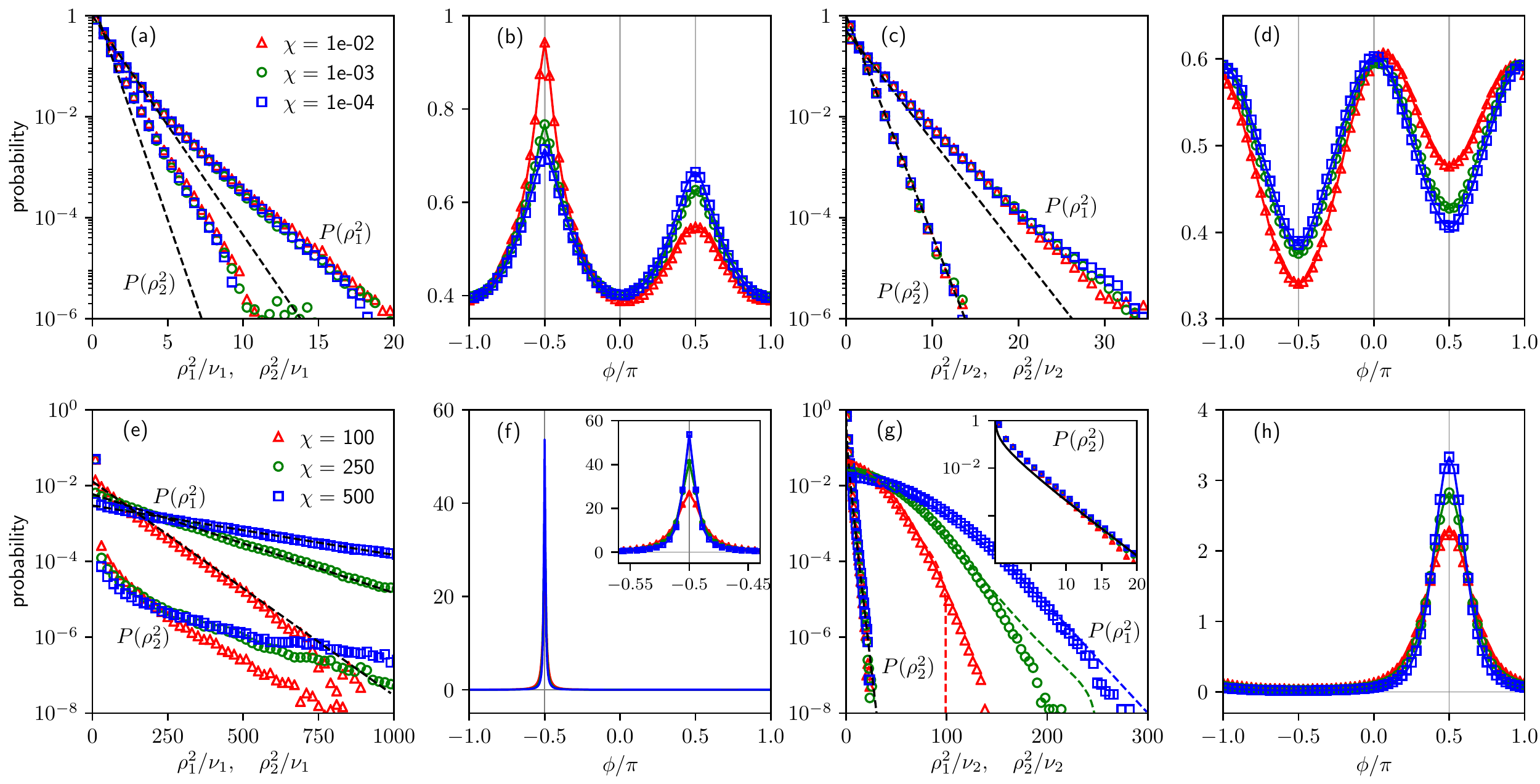}
	\caption{    Probabilities of the occupation numbers and the phase for:
		(a, b) inverse cascade, small $\chi$;
		(c, d) direct cascade, small $\chi$;
		(e, f) inverse cascade, large $\chi$;
		(g, h) direct cascade, large $\chi$.
		Each dataset contains 20M datapoints, at $\Delta t = 0.01$ for inverse cascade and large $\chi$
		and  $\Delta t = 0.1$ for all other cases. For the inverse cascade, $\chi = \gamma_1^3 / (2 P_2 V^2)$ and $\nu_1 = P_2/\gamma_1$.
		We use $\gamma_{1,2} = 0.01$  for small $\chi$ and $\gamma_{1,2} = 1$ for large $\chi$. In all cases, $V=1$. Broken lines in (g) correspond to the approximation  (\ref{invweakP2},\ref{invweakP1}).
	}
	\label{ICDCprob}
\end{figure*} 

\subsection{Small Reynolds number: strong-interaction-weak-noise limit}
\label{smallRe}


\subsubsection{Inverse cascade}

In an inverse cascade, energy goes from high to low frequency, so we set $\gamma_2=0$ and $P_1=0$ in (\ref{gen00},\ref{gen0}):
\bea
\label{inv1}
&&\dot b_1=-2i V^*b_1^*b_2-\gamma b_1,\\
\label{inv1_0}
&&\dot b_2=-i Vb_1^2+\xi(t).
\eea
In the steady state, the energy input rate $P$ must be equal to the dissipation rate $\gamma n_1$ and to the energy flux from the second mode to the first given by the imaginary part of the third cumulant: $2V\langle \rho_1^2\rho_2\sin\theta\rangle = -P$. Then, from the energy balance we obtain $n_1=P/\gamma$ and ${2V}\langle \rho_1^2\rho_2\sin\theta\rangle =-{P}$,
	so that $\langle \rho_1^2\rho_2\rangle \ge {P}/{2V}$.
	Also, from the condition $\frac{d}{dt}\langle\ln |b_1|^2\rangle=0$ we find
	$2V\langle{\rho_2}\sin\theta\rangle=- {\gamma}$
	and, therefore $\langle{\rho_2}\rangle\ge{\gamma}/ {2V} $.


From  Eqs. (\ref{inv1},\ref{inv1_0}) it is straightforward to see that when $\chi\ll 1$
the steady-state probability distribution ${\cal P}(b_1,b_2)$ cannot be close to the equilibrium Gaussian (\ref{SG2}) with the temperature $T=2P/\gamma$ and the equipartition $P/\gamma=n_{1}=2n_{2}$.
Indeed,  the stationarity of $ \langle {\cal H}^2\rangle=4|V|^2\langle\rho_1^4\rho_2^2\cos^2\theta\rangle$ requires $\langle {\cal H}^2\rangle/\langle \rho_1^4\rangle= \frac{|V|^2P}{2\gamma}$, but
this contradicts the Gaussian ratio which gives $\langle {\cal H}^2\rangle/\langle \rho_1^4\rangle=\frac{\left|V\right|^{2}P}{\gamma}$.


Thus, small value of $\chi$ does not mean that the system is near equipartition. 
In agreement with this conclusion, results of numerical modeling presented at two lower left panels  of Figure \ref{ICDCprob} show that at  neither of marginal distributions of the mode amplitudes is close to Gaussian and that the phase distribution does not become uniform as $\chi$ tends to zero.
This is also reflected in nonzero value $I_{12}(+0)$ of mutual information between modes in this limit, 
see the right panel of Figure~\ref{tm_mi_ext}.

\subsubsection{Direct cascade}\label{sec:smalldir}
Direct cascade corresponds to the choice $\gamma_1=0$ and $P_2=0$ in  (\ref{gen00},\ref{gen0}):
\bea
\label{bb1_0}
&&\dot b_1=-2i V^*b_1^*b_2+\xi(t),\\
&&\dot b_2=-i Vb_1^2-\gamma b_2.
\label{bb1}\eea
Now energy goes from low to high frequency.
Again, in the steady-state regime, the energy input rate $P/2$ must be equal to the dissipation rate $2\gamma n_2$ and to the energy flux from the first mode to the second given by the imaginary part of the third cumulant:
	$2V\langle \rho_1^2\rho_2\sin\theta\rangle = P/2$, so $n_2=P/4\gamma$.
	From $\frac{d}{dt}\langle\ln |b_2|^2\rangle=0$ we find
	$\langle \rho_1^2 \sin\theta /\rho_2\rangle= {\gamma}/{V}$. Therefore  $\langle \rho_1^2\rho_2\rangle \ge{P}/{4V}$ and $\langle{\rho_1^2}/{\rho_2}\rangle\ge{\gamma}/{V}$.

When $\chi\to0$, the dimensionless flux $ \langle \rho_1^2\rho_2\sin\theta\rangle/n_1n_2^{1/2}=\chi^{1/2}$ is small, which may suggest that phase-space distribution is close to the Gaussian equilibrium (\ref{SG2}) with $T=P/\gamma$
and that the phase distribution is close to uniform. Furthermore, as opposed to 
the case of inverse cascade discussed above, the equality obtained from the
stationarity of $\langle {\cal H}^2\rangle$:
\be  \langle {\cal H}^2\rangle=4|V|^2\langle\rho_1^4\rho_2^2\cos^2\theta\rangle=\frac{2|V|^2 P}{\gamma}\langle \rho_1^2\rho_2^2\rangle\ ,\label{K2dir}\ee
is achieved by the Gaussian distribution with $n_1=2n_2=P/2\gamma$. 
However, numerical data, as can be seen from two upper left panels of Figure~\ref{ICDCprob}, shows that even though the marginal  distributions of amplitudes are close to Gaussian with equipartition, $n_1\approx2n_2$, the phase distribution is far from flat and deviation from equilibrium is substantial. 
The mutual information between modes as a function of $\chi$ exhibits a non-zero value of $I_{12}(+0)$ (see Figure~\ref{tm_mi_ext}) which is also a clear footprint of non-equilibrium.

\subsection{Large Reynolds number: weak interaction, strong noise limit}
\label{sec:largeRe}

\subsubsection{Inverse cascade}\label{sec:inv}

The pair of complex equations (\ref{inv1},\ref{inv1_0}) can be rewritten as three real ones since the overall phase drops out:
\bea
\label{inv001}
&& \dot \rho_1= - 2|V|\rho_1\rho_2\sin\theta-\gamma \rho_1,\\
\label{inv002}
&&\dot \rho_2= |V|\rho_1^2\sin\theta+{P\over4\rho_2}+{\zeta(t)\over\sqrt2},\\
\label{inv003}
&&\dot\theta=|V|\frac{\rho_1^2-4\rho_2^2}{\rho_2}\cos\theta+{\frac{\zeta(t)}{\sqrt{2}\rho_2}},
\label{inv}\eea
 where $\zeta(t)$ is the real white noise with zero mean $\langle \zeta(t)\rangle=0$ and the pair correlation function   $\langle \zeta(t_1)\zeta(t_2)\rangle=P\delta(t_1-t_2)$.

When $\chi\gg 1$, Eqs. (\ref{inv001})-(\ref{inv003}), can be further simplified by assuming that relative phase is locked on $\theta=-\pi/2$ most the time.
Then, one  gets the following closed equations for the amplitudes dynamics
\begin{eqnarray}
\label{inv200}
&&\dot \rho_1=  2|V|\rho_1\rho_2-\gamma \rho_1,\\
&&\dot \rho_2=-|V|\rho_1^2+{P\over4\rho_2}+{\zeta(t)\over\sqrt2}.
\label{inv2}
\end{eqnarray}
A hypothesis that the modes are statistically  independent in this limit is shown incorrect in the Appendix~\ref{sec:ap2}. This result is in sharp contrast with the model described in \cite{Courant1}, where
	authors found the factorized joint probability density ${\cal P}(\rho_1,\rho_2)$ of mode amplitudes  in the limit when their analogue of the parameter $\chi$ is large.

While constructing the probability densities for inverse cascade at $\chi\to\infty$ turns out to be a tricky task, it is straightforward to describe general features of stochastic dynamics dictated by Eqs. (\ref{inv200}) and (\ref{inv2}).
Namely, these  pair of nonlinearly coupled equations suggest the following cyclical evolution: $\rho_1$ stays close to zero most of the time while $\rho_2$  undergoes diffusion in a repulsive logarithmic potential; when $\rho_2$ sufficiently outgrows the threshold level $\gamma/2|V|$, $\rho_1$ shoots up and quickly diminishes $\rho_2$; after that $\rho_1$ also resets to the near-zero level and the stochastic dynamics of $\rho_2$ starts from scratch.
The mode dynamics during the intermittent burst events can be described by simplified equations
\begin{eqnarray}
\label{noise_free1}
\dot{\rho}_1=2|V|\rho_1\rho_2-\gamma\rho_1,\\
\label{noise_free2}
\dot{\rho}_2=-|V|\rho_1^2.
\end{eqnarray}
Compared with Eqs. (\ref{inv200}) and (\ref{inv2}),  we neglected the terms associated with noise.
\begin{figure}[t] 
	\centerline{\includegraphics[width=60mm]{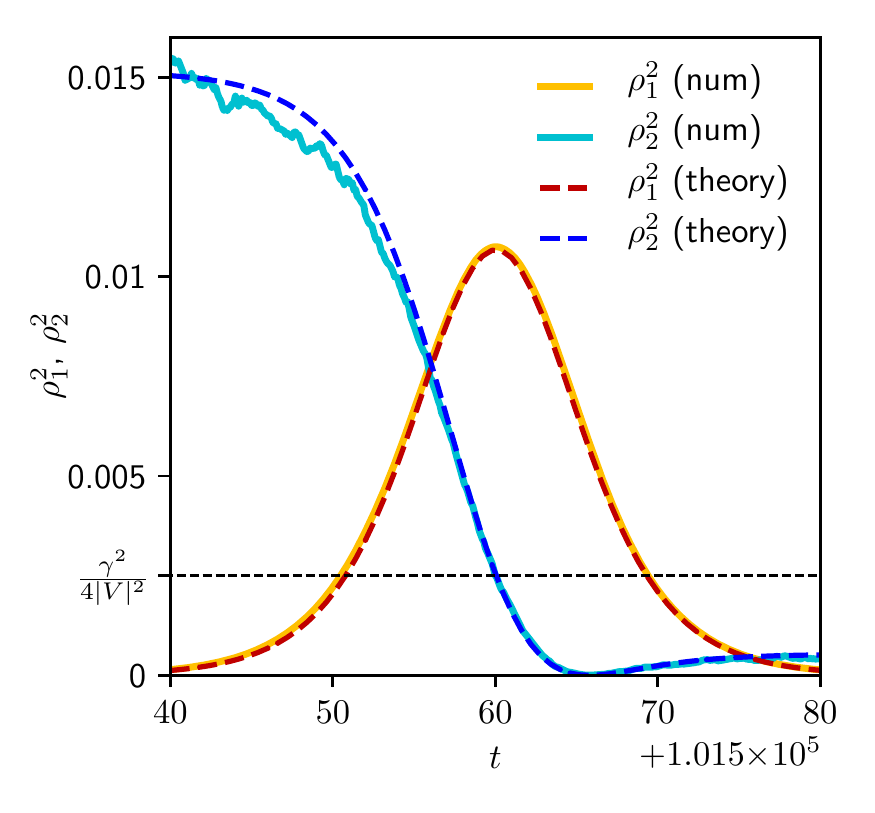}}
	\caption{An individual realization
of the modes trajectories during one of the events. The continues lines are obtained from numerical simulations of Eqs.  (\ref{inv1}) and (\ref{inv1_0}, and the dashed lines represent theoretical fit based on Eqs. (\ref{burst_rho1}) and (\ref{burst_rho2})). }
	\label{pic:burst}
\end{figure}
Equations (\ref{noise_free1}) and (\ref{noise_free2}) are exactly solvable yielding
\be  {\rho}_1^2(t)= {r_1^2-2\left(\rho_2(t)-\frac{\gamma}{2 V }\right)^2+2\left(r_2-\frac{\gamma}{2|V|}\right)^2}\,,\label{burst_rho1}\ee
\begin{widetext}
	\begin{eqnarray}
	\label{burst_rho2}
	&&{\rho}_2(t)=\frac{\gamma}{2V}+\sqrt{\left(r_2-\frac{\gamma}{2V}\right)^2+\frac{r^2_1}{2}}  \tanh\left[\frac12\ln\frac{\sqrt{\bigl(r_2-
			\frac{\gamma}{2V}\bigr)^2+\frac{r^2_1}{2}}+r_2-\frac{\gamma}{2V}}{\sqrt{\bigl(r_2-
			\frac{\gamma}{2V}\bigr)^2+\frac{r^2_1}{2}}-r_2+\frac{\gamma}{2V}} 
	-2Vt\sqrt{\Bigl(r_2-\frac{\gamma}{2V}\Bigr)^2+\frac{r^2_1}{2}} \right],
	\end{eqnarray}
\end{widetext}
where $r_1=\rho_1(0)$, $r_2=\rho_2(0)$ are the initial conditions.
Estimating $r_2\sim  {\gamma/V}$ and $r_1\ll r_2$, we  see from Eq. (\ref{burst_rho2})  that the duration of such burst event is $\sim \gamma^{-1}$, which is much smaller than the typical inter-events period $\sim  {\gamma^2}/{PV^2}$.
As follows from Eq. (\ref{burst_rho1}), the amplitude of the dissipated mode grows from the initial value $r_1\ll r_2$ to the maximum value $\rho_{1\text{max}}=\sqrt{r_1^2+2(r_2-\frac{\gamma}{2V})^2}\approx \sqrt2(r_2-\frac{\gamma}{2V})$ (attaining it at the moment when $\rho_2(t)=\frac{\gamma}{2V}$) and finally returns to the starting level $\rho_1(\infty)=r_1$. Such bursts are likely responsible for pulses running in shell models, which are chains of interacting triplets.

Numerical simulations confirm the intermittent nature of system dynamics described above.  
Namelly, panel (f) of Figure \ref{ICDCprob} reveals that the relative phase is indeed locked at $-\pi/2$.
Figure \ref{pic:burst} illustrates that analytical prediction based on the assumption of phase locking (see Eqs. (\ref{burst_rho1}) and (\ref{burst_rho2})) are in excellent agreement with numerical data extracted from simulations  of Eqs.  (\ref{inv1}) and (\ref{inv1_0}).
As visible in panel (e) of Figure \ref{ICDCprob}, the tails of the amplitudes probability distributions strongly depend on $\chi$; the fits of ${\cal P}(\rho_1)$ and ${\cal P}(\rho_2)$ by the broken lines in Figure \ref{ICDCprob} are empirical.
As for the mutual information, 
from Figure~\ref{tm_mi_ext} we see that  $I_{12}\propto \ln\chi$ for inverse cascade with $\chi\gg1$.

\subsubsection{Direct cascade}

We conclude the treatment of our turbulent cascades with the case of a direct cascade in the limit of large Reynolds number, $\chi\rightarrow\infty$. It is in this limit we find a window to the way a singular measure is formed far away from equilibrium. We indeed find that in this limit the full probability distribution is singular and thus corresponds to the lowest entropy state $S_{12}\rightarrow-\infty$.

\label{sec:dir}
In addressing the weak-interaction-strong-noise limit $\gamma\to\infty$ ($\chi\rightarrow\infty$), it is convenient to
express $b_2$ from  (\ref{bb1}) as an integral, which
in the leading order shows that the second amplitude is enslaved to the first one:
\begin{equation}
\label{bb3}
b_2(t)=-iV\int_{-\infty}^tb_1^2(t')e^{\gamma(t'-t)}dt'\approx -\frac{iVb_1^2}{\gamma}.
\end{equation}
Substituting  this relation into (\ref{bb1_0}), we get a closed equation on the stochastic dynamics of $b_1$
\be \dot b_1= -\frac{2V^2}{\gamma}b_1^*b_1^2+\xi(t),\label{SG4}\ee
from which one finds the following expressions for the marginal probability distributions
\begin{align}
{\cal P}(b_1)&=Z_1^{-1}\exp\biggl(-\frac{2 V^2}{\gamma P}|b_1|^4\biggr),
\label{invweakP1_01}\\
{\cal P}(b_2)&=Z_2^{-1}\exp\biggl(-\frac{2\gamma}{P}|b_2|^2\biggr),
\label{invweakP2}
\end{align}
which are valid at $\rho_1,\rho_2\ll {\gamma}/{|V|}$. Thus, the whole probability density in the four-dimensional phase space is  singular at $\chi\to\infty$, sitting on a three-dimensional manifold
\be\!\!\! {\cal P}(b_1,b_2)=\frac{2}{\sqrt{\pi P\gamma}}\exp\biggl(-\frac{4V^2|b_1|^4}{P\gamma}\biggr)\delta(b_2-\frac{iV}{\gamma}b_1^2)\,,\label{SG5}\ee
so that the total entropy  $S_{12}\to-\infty$. 
Note also that Eq. (\ref{SG5}) yields  large ratio of the typical mode amplitudes: $\rho_1^2/\rho_2^2\simeq \sqrt{\gamma^3/P|V|^2}=\sqrt\chi\gg1$.

Since the distribution over the overall phase is flat, one can integrate it out and conclude that the distribution in the the three-dimensional space of variables $\rho_1,\rho_2,\theta$ concentrates on the 
curve $\rho_2\propto\rho_1^2$.
Interestingly,  with increasing $\chi$ the joint probability distribution ${\cal P}(b_1,b_2)$ is getting sharper than Gaussian along this curve. This is different from the model discussed in  \cite{Courant1} where the driving mode  is nearly Gaussian in this limit, the relative amplitudes of both modes fluctuate, and only the relative phase is fixed, that is  the joint probability density is only singular with respect to the phase difference $\theta$.

Considering large but finite $\chi$, instead of delta-function in Eq. (\ref{SG5}) one obtains the distribution with a finite width which is the variance of the difference $b_2-Vb_1^2/\gamma$.
To estimate this width we further expand Eq. (\ref{bb3})
\begin{align}
\label{bb3_0}
b_2(t)\approx&-i V\int_{-\infty}^t\left[b_1^2(t)+(t'-t){d b_1^2(t)\over dt}\right]e^{\gamma(t'-t)}dt'\\
=&-{i Vb_1^2(t)\over \gamma}+{2i Vb_1(t) \over \gamma^2}{db_1(t)\over dt}.
\end{align}
From Eq.  (\ref{bb3_0}) we get 
\be\langle |b_2+iVb_1^2/\gamma|^2\rangle={4|V|^2\over \gamma^4}\langle| b_1\dot b_1|^2\rangle= {4|V|^2P^2 \over \gamma^4}\ .\label{width}\ee
Dividing this result by 
$\langle |b_2|^2\rangle=P/\gamma$,  one  obtains that the relative squared width behaves as $1/\chi$
(as expected, it tends to zero when $\chi\to\infty$).
The non-zero width at finite values of $\chi$  entails the finite entropy of the distribution ${\cal P}(b_1,b_2)$: $S_{12}\simeq -\ln \chi$.
This analytical prediction  is supported by  Figure~\ref{tm_mi_ext}.
For comparison, the naive Gaussian ansatz yields $S_1+S_2\simeq \ln n_1n_2\propto  \ln \chi^{-1/2}$, since $n_2=P/4\gamma$ and $n_1\simeq \sqrt{P\gamma}/|V|$.

Now let us plug Eq. (\ref{bb3}) into Eq.(\ref{bb1}), then solving the corresponding steady Fokker-Planck equation (see \ref{sec:apLD}) one arrives at the next order correction for the marginal probability distribution for $\rho_1\ll {\gamma}/{|V|}$:
\begin{equation}
{\cal P}(\rho_1)\propto \rho_1(1-\frac{4|V|^2}{\gamma^2}\rho_1^2)\exp(-\frac{2|V|^2}{\gamma P}\rho_1^4+\frac{16|V|^4}{3\gamma^3P}\rho_1^6)\,.
\label{invweakP1}
\end{equation}
which is more accurate than Eq. (\ref{invweakP1_01}). Unfortunately, extracting similar correction for the probability density ${\cal P}(\rho_2)$ as well as the further high order corrections to ${\cal P}(\rho_1)$ is more challenging. 

As can be seen from panel (g) of Figure \ref{ICDCprob},  Eqs. (\ref{invweakP2}) and (\ref{invweakP1})  allow us to fit the numerical data  quite accurately in the range  $\rho_1,\rho_2\ll {\gamma}/{|V|}$.
Expectedly, the agreement between numeric and analytic results improves with the growth of $\chi$.


\section{Laser generation}\label{sec:laser}

Let us now pump  the first harmonic by an instability, for instance, in a laser. Consider first pure dynamics, setting $\xi_1=\xi_2=0$ in (\ref{gen00},\ref{gen0}) and changing sign in front of $\gamma_1$, which now describes gain for an optical signal.  
Then the resulting evolution satisfies three closed equations:
\begin{eqnarray}
\label{SG600}
&&{d\rho_1^2 \over dt}=-2F+2\gamma_1\rho_1^2,\\
\label{SG60}
&&{d\rho_2^2\over dt}=F-2\gamma_2\rho_2^2,\\
\label{SG6}
&&  {dF\over dt}=(2\gamma_1-\gamma_2)F+2|V|^2\bigl( \rho_1^4-4\rho_1^2\rho_2^2\bigr),
\end{eqnarray}
where $F=2V\rho_1^2\rho_2\sin\theta$ is the instantaneous flux.

Apart from the trivial unstable fixed point $\rho_1=\rho_2=0$, Eqs. (\ref{SG600})-(\ref{SG6}) have the stationary point $\bar\rho_1^2=\gamma_1\gamma_2/2|V|^2$, $\bar\rho_2^2=\gamma_1^2/4|V|^2$,  $ \bar \theta=\pi/2$ (and, thus, $\bar F=\gamma_2\gamma_1^2/2|V|^2$). 
This means that in the the degenerate case $\gamma_2=2\gamma_1$, the system possesses the steady state $\rho_1=2\rho_2$ existing for any $\theta$. This marginal stability turns into an instability of the steady state at $\gamma_2<2\gamma_1$ and into stability at $\gamma_2>2\gamma_1$.
In what follows, we consider  $\gamma_2>2\gamma_1$.
Note that in practice both $\gamma_1$ and $\gamma_2$ often depend on the amplitudes, for instance, due to gain saturation or/and nonlinear damping. However, our main focus here is on the noise impact on the steady state, so we will treat $\gamma_1,\gamma_2$ taken near this state as constants.

{
	Let us now add a random pumping  and study its influence on the efficiency and statistics of conversion.
	The modes amplitudes, $\rho_1$ and $\rho_2$ and the relative phase $\theta$ are governed by the following equations
	\begin{eqnarray}
	&&\dot \rho_1= - 2V\rho_1\rho_2\sin\theta+\gamma_1\rho_1 +{P\over4\rho_1}+\frac{\zeta_1(t)}{\sqrt{2}},\\
	\label{laser2}
	&&\dot \rho_2= V\rho_1^2\sin\theta-\gamma_2 \rho_2\,,\\
	&& \dot\theta={\rho_1^2-4\rho_2^2\over\rho_2}V\cos\theta+{\sqrt2\zeta_2(t)\over\rho_1}.
		\label{laser2'}
	\end{eqnarray}
	Here $\zeta_1$ and $\zeta_2$ are two independent real white noises with zero mean values $\langle \zeta_i(t)\rangle=0$, and the pair correlator  $\langle \zeta_i(t_1)\zeta_j(t_2)\rangle=P\delta_{ij}\delta(t_1-t_2)$.
	In the limit of weak noise, $P|V|^2\ll\gamma_1^2(\gamma_2-2\gamma_1)$, one can 
	apply a linear approximation near the fixed point.
	More specifically, we substitute  decomposition $\rho_1(t)=\bar\rho_1+u(t),\rho_2(t)=\bar\rho_2+v(t), \theta(t)=\bar \theta+\phi(t)$ into Eqs. (\ref{laser2})-(\ref{laser2'}) and keep only the first order terms with respect to $u$, $v$ and $\phi$.
	This procedure yields 
	
		\begin{eqnarray}
	\label{laser3'}
	&&\dot u=  - 2V\bar\rho_1v +{P\over4\bar\rho_1}+\frac{\zeta_1(t)}{\sqrt{2}},\\
	&&\dot v= 2V\bar\rho_1u-\gamma_2 v,\\
	&&\dot\phi={(2\gamma_{1}-\gamma_{2})}\phi+{\sqrt{2}\zeta_2(t)\over\bar\rho_1}.\label{laser3}
	\end{eqnarray}
From Eqs. (\ref{laser3'})-(\ref{laser3}) we immediately find the
 variances 


		\begin{eqnarray}
	&& \langle u^2\rangle- \langle u\rangle^2=\frac{\left(2\gamma_{1}+\gamma_{2}\right)P}{8\gamma_1\gamma_2}\,,\\
	&&
	\langle v^2\rangle-\left\langle v\right\rangle ^{2}=\frac{P}{4\gamma_{2}}\,,\\
	&&\langle \phi^2\rangle-\langle \phi\rangle^2=\frac{2V^{2}P}{\gamma_{1}\gamma_{2}\left(\gamma_{2}-2\gamma_{1}\right)}.
	\end{eqnarray}
We see that level of fluctuations in relative phase grows when one approaches the stability threshold. Note also that far from the  threshold, i.e. at $2\gamma_1\ll\gamma_2$,  the fluctuations of the second harmonic are suppressed: $\langle v^2\rangle/\langle u^2\rangle\approx \gamma_1/\gamma_2\ll1$. In this case, the noise of the first harmonic only weakly influences the conversion into the second one. However, the conversion is least effective in this limit: $\bar\rho_2^2/\bar\rho_1^2=\gamma_1/2\gamma_2\ll1$.



\section{Conclusion}
\label{sec:con}

Our most important finding is the explicit formula (\ref{SG5}) for the singular measure of a direct cascade in the limit of strong noise and weak interaction. We believe that this is a meaningful advance in non-equilibrium statistics, as it opens a window to the study of the formation of singular measures in systems driven far away from equilibrium. We have described also the approach to this limit and have shown that the total entropy decays and the inter-mode mutual information grows logarithmically with the Reynolds number. In the inverse cascade case in this limit, the phase is locked on $-\pi/2$ and the system exhibits an intermittent dynamics of bursts, which we were able to describe analytically. Such bursts are perhaps responsible for pulses running in shell models, which are chains of interacting triplets used in modeling hydrodynamic incompressible turbulence. It is thus may be interesting to apply the methods developed here to the popular shell model $\dot u_i=u_{i-1}^2-u_iu_{i+1}$ \cite{DN}. After some elementary transformations, this model can be turned into that with the Hamiltonian ${\cal H}=\sum_iV_i(a_i^2a_{i+1}^*+c.c)$, that is the interacting chain built of our pairs.

The opposite limit of weak noise and strong interaction is a singular one: the probability distribution is not close to a Gaussian distribution determined by $N$ for however small $\chi$, despite occupation numbers being close to equipartition and the marginal one mode distributions close to quadratic. It is expressed, in particular, in the nonzero mutual information $I_{12}(\chi)$ at the limit $\chi\to+0$. Figure~\ref{tm_mi_ext} combines the mutual information data for both cascades. We see that  $I_{12}(\chi)-I_{12}(+0)\propto \chi^2$ at $\chi\ll1$.  We failed to find an analytic solution in this limit either in the direct or inverse cascade, even though it is likely that the probability distribution can be expressed in terms of $N $ and ${\cal H}^2$, which are the conserved quantities of the unforced undamped system.


We conclude with suggesting an interesting application of our model to wave turbulence. In a set of $M+2$ interacting waves, one may consider to model the interaction of a resonant couple with the other $M$ waves by dissipation and random forcing.
When $M\gg1$ we can treat forces from all other modes as a white noise, so that our model (\ref{gen0}) applies. In this case, different limits in $\chi$ correspond to different situations. If we assume an almost continuous distribution of other modes and estimate from the wave kinetic equation $\gamma\simeq V^2Mn/\omega$ and $P\simeq \gamma n$ \cite{ZLF}, then $\chi=\gamma^3/PV^2\simeq V^2Mn/\omega^2\ll1$, which is the original parameter of nonlinearity assumed to be small. In this case, we come to the surprising conclusion that a resonant mode within turbulence, when $\Delta T\simeq T$, has a relative entropy of order unity and independent of $V$. If, however, we have a set of well-isolated resonant interactions, then it  makes more sense to assume that the interaction with a given mode is $M$ times smaller so that $\chi$ is large (as $M$ or $\sqrt{M}$), then the relative entropy is small. Note that in most cases the number of resonant interactions, 
is much less than the total number of modes in the system.

The work was supported by the Scientific Excellence Center and Ariane de Rothschild Women Doctoral Program at WIS, grant  662962 of the Simons  foundation, grant 075-15-2019-1893 by the Russian Ministry of Science,  grant 873028 of the EU Horizon 2020 programme, and grants of ISF, BSF and Minerva. NV was in part supported by NSF grant number DMS-1814619. This work used the Extreme Science and Engineering Discovery Environment (XSEDE), which is supported by NSF grant number ACI-1548562, allocation DMS-140028.

	\nocite{*}
	\bibliographystyle{unsrt}
	\bibliography{aps2}

	\appendix
	\section{Appendix}
	
	\subsection{Hamiltonian evolution}\label{sec:Ham}

	
	Here we briefly sketch some of the elementary properties of the Hamiltonian system defined by Eq. (\ref{Ham5}).
	From  Eq. (\ref{Ham5}) we find the system of two coupled complex equations
	\begin{eqnarray}
	\label{HamEq1}
	&&\dot a_1=-i\frac{\partial {\cal H}}{\partial a_1^*}= -i\omega a_1-2iVa_1^{*}a_2,\\
	\label{HamEq2}
	&&\dot a_2=-i\frac{\partial {\cal H}}{\partial a_2^*}=-2i\omega a_2-iV^{*}a_1^2.
	\end{eqnarray}
	It is easy to see that in addition to ${\cal H}$, Eqs. (\ref{HamEq1}) and (\ref{HamEq2}) have the second integral of motion,
	\begin{equation}
	\label{N}
	N=\omega|a_1|^2+2\omega|a_2|^2,
	\end{equation}
	and, thus, the system is completely integrable.
	Indeed, two integrals of motion allows one to reduce Eqs.  (\ref{HamEq1}) and (\ref{HamEq2})  to a single first-order equation, which we write for $x=2|a_2|^2/N$ and $t\to tVN^{1/2}$
	\be {dx\over dt}=\pm 2\sqrt{2x(1-x)^2-4K^2/N^{3}}\ ,\label{Ham6}\ee
	where
	\begin{equation}
	\label{K}
	K={\cal H}-N=V  a_1^{*2}a_2+  V^*a_1^{ 2}a_2^*,
	\end{equation}
	is also an integral of motion.
	
	\begin{figure}[h]
		\begin{singlespace}
			\includegraphics[scale=0.21]{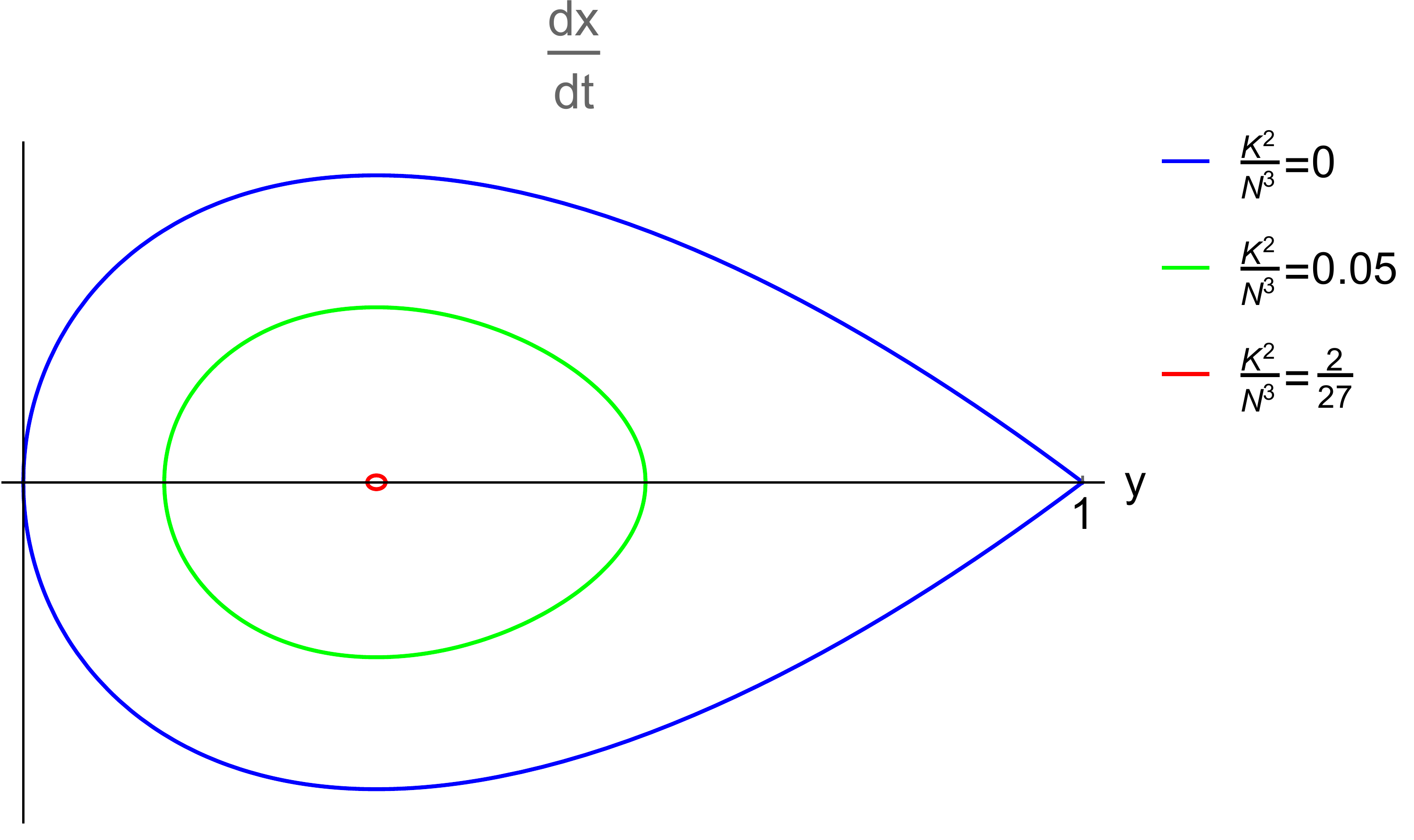}
		\end{singlespace}
		\caption{
			\label{fig:tori} The phase portrait of the integrable Hamiltonian dynamics plotted for different values of the dimensionless ratio of integrals of motion, $K^2/N^3$.
		}
	\end{figure}

	Next, using the Euler representation, $a_1=\rho_1e^{i\varphi_1}$ and  $a_2=\rho_2e^{i\varphi_2}$, one obtains from Eqs. (\ref{HamEq1}) and (\ref{HamEq2})
	\begin{eqnarray}
	&&\dot \rho_1= - 2|V|\rho_1\rho_2\sin\theta\,,\\
	&&\dot \rho_2= |V|\rho_1^2\sin\theta\,,\\
	&&\dot\theta={\rho_1^2-4\rho_2^2\over\rho_2}\cos\theta,\label{rho0}
	\end{eqnarray}
	where $\theta= \arg a_1^2a_2^*=2\phi_1-\phi_2$.
	From Eqs. (\ref{Ham6}) - (\ref{rho0}) one finds that the dynamical system has two fixed points: 1)  $x=1=2\rho_2^2/N$,  $K=0,$ which means   $\rho_1=0$ and $\theta= \pm\pi/2$, and 2) $x=1/3$, $K^2/N^3=2/27$, which means $\rho_1^2=4\rho_2^2=2N/3$ and $\theta=0$ or $\theta=\pi$.
	It is easy to show that the first one is unstable, while the second one is stable.
	The small oscillations near the second point are harmonic with the frequency $4\sqrt2/3$.
	We will see below that this phase portrait explains qualitatively the statistics of system in the presence of dumping and (small) random forcing.
	
	Since $K=2V\rho_1^2\rho_2\cos\theta$ is a constant, it cannot change sign, an, therefore,  we have two separated regions in phase space that correspond to the sign of $\cos\theta$: $\left[-\pi/2,\pi/2\right]$ and $\left[\pi/2,3\pi/2\right]$.
	Both separating planes at $\theta=\pi/2,3\pi/2$ are critical points of $\phi$, on which non-linear interaction is zero.
	
    \subsection{Small-flux limit}\label{sec:smallflux}
    
    This limit can be called alternatively either small-flux limit, because the mean value of the inter-mode energy flux $ 2\langle\rho_1^2\rho_2\sin\theta\rangle$ is much smaller than its periodic oscillation, or small-noise limit, because most of the time the evolution is unaffected by pumping and damping. The phase portrait described in Section~\ref{sec:Ham} explains qualitatively the statistics of turbulence in this limit of large wave amplitudes, presented in the upper row of Figure \ref{ICDCprob}. Indeed, the dynamical system has two fixed points: the unstable one $\rho_1=0$, $\theta= \pm\pi/2$, and the stable one $\rho_1^2=4\rho$, $\theta=0,\pi$. In this limit, the system spends most of its time close to one of the fixed points, randomly switching between them. In the direct cascade, described in Sect~\ref{sec:smalldir}, random noise acts on the first mode, so that the system spends less time around $\rho_1$ close to zero, and the probability has minima at $\theta= \pi/2,-\pi/2$. In the direct cascade, the system spends more time oscillating around the second fixed point and the probability has maxima at $\theta=0,\pi$. On the contrary, dissipation acts on $\rho_1$ in the inverse cascade, which keeps system longer around the first fixed point, and the probability has maxima at $\theta= \pi/2,-\pi/2$.

	\subsection{Large-flux limit for the direct cascade}\label{sec:apLD}
	
	Here we briefly discuss the derivation of the solution Eq. (\ref{invweakP1}) for the marginal probability distribution for $\rho_1$ in the main text. 
	
	Let us plug Eq. (\ref{bb3}) into Eq. (\ref{bb1}) to get:
	\begin{equation}
	\dot b_1=-\frac{2|V|^2}{\gamma}\frac{|b_1|^2b_1}{1-\frac{4|V|^2}{\gamma^2}|b_1|^2}+
	\frac{\xi(t)}{1-\frac{4|V|^2}{\gamma^2}|b_1|^2}.
	\end{equation}
	The steady-state probability density of the amplitude $\rho_1$ obeys the  Fokker-Planck equation
	\begin{eqnarray}
	&&\frac{2|V|^2}{\gamma}\frac{1}{\rho_1}\frac{\partial}{\partial \rho_1}\left[\frac{\rho_1^3}{1-\frac{4|V|^2}{\gamma^2}\rho_1^2}{\cal P}(\rho_1)\right] +\\
	&&+\frac{P}{4\rho_1}\frac{\partial}{\partial \rho_1} \left[\rho_1 \frac{\partial}{\partial \rho_1}\left[\frac{\rho_1^{-1}}{(1-\frac{4|V|^2}{\gamma^2}\rho_1^2)^2}{\cal P}(\rho_1)\right]\right]=0.
	\end{eqnarray}
	Solving this equation one arrives at Eq. (\ref{invweakP1}) in the main text.

	\subsection{Large-flux limit for the inverse cascade}\label{sec:ap2}
	The stationary Fokker-Planck equation on ${\cal P}(\rho_1,\rho_2)$ is as follows
	\be  {P\over4}{\partial^2 {\cal P}\over\partial \rho_2^2}+{\partial\over\partial \rho_2}\biggl[|V|\rho_1^2-{ P\over4\rho_2}\biggr]{\cal P}+{\partial\over\partial \rho_1}\biggl[\gamma\rho_1- 2|V|\rho_1\rho_2\biggr]{\cal P}
	=0\ .\label{inv3}\ee
	
	From  Eq. (\ref{inv2}) we obtain
	\begin{eqnarray}
	\label{inv6}
	&&\langle\rho_2\rangle=\frac{\gamma}{2|V|},\\
	\label{inv7}
	&&\langle \rho^n_1\rangle=\frac{2|V|}{\gamma}\langle \rho_1^n\rho_2\rangle,\\
	\label{inv7.5}
	&&\langle \rho_1^2\rangle=\frac{P}{4|V|}\langle \frac{1}{\rho_2}\rangle,
	\end{eqnarray}
	Multiplying FPE (\ref{inv3}) by $\rho_2^2$ and integrating over $d\rho_1d\rho_2$
	yields
	\begin{equation}
	\label{inv8}
	\langle\rho_1^2\rho_2\rangle=\frac{P}{2|V|},
	\end{equation}
	and therefore (due to Eqs. (\ref{inv7}),  (\ref{inv7.5}) and (\ref{inv8}))
	\begin{eqnarray}
	&&\label{inv9}
	\langle \rho_1^2\rangle=\frac{P}{\gamma},\\
	\label{inv010}
	&&\langle \frac{1}{\rho_2}\rangle=\frac{4|V|}{\gamma}.
	\end{eqnarray}
	Also from Eq. (\ref{inv3}) we find
	\begin{equation}
	\langle\rho_2^n \rangle=\frac{4|V|}{(n+2)P}\langle \rho_1^2\rho_2^{n+1}\rangle,
	\quad
	\langle \rho_1^2\rho_2^{2} \rangle=\frac{3}{8}\frac{\gamma P}{|V|^2}.
	\end{equation}
	It follows from Eqs. (\ref{inv6}), (\ref{inv7}), (\ref{inv8}) and (\ref{inv9}) that
	\begin{equation}
	\langle \rho_1\rho_2\rangle=\langle \rho_1\rangle\langle\rho_2\rangle, \ \ \ \langle \rho_1^2\rho_2\rangle=\langle \rho_1^2\rangle\langle\rho_2\rangle,
	\end{equation}
	which may lead one to hypothesize that in the steady state the random variables $\rho_1$ and $\rho_2$ are statistically independent. If such statistical independence was true, then
	\begin{equation}
	{\cal P}(\rho_1,\rho_2)=C \rho_1^{-1+\frac{4P|V|^2}{\gamma^3}}e^{-\frac{2|V|^2\rho_1^2}{\gamma^2}}\rho_2 e^{-\frac{4|V|}{\gamma}\rho_2}.
	\label{invP12}
	\end{equation}
	However, direct substitution of Eq. (\ref{invP12}) into the Fokker-Planck equation (\ref{inv3}) shows that this distribution represents the solution only along two lines: $\rho_2=(\frac{1}{2}+\frac{1}{2\sqrt{2}})\frac{\gamma}{|V|}$ and $\rho_1=\sqrt{\frac{P}{\gamma}}$.
	Thus, the hypothesis of statistical independence is not self-consistent.
	
	\end{document}